\documentclass[aps,prb,showkeys,twocolumn,showpacs,floatfix,10pt]{revtex4-1}
\usepackage{graphics,graphicx}
\usepackage{color}
\usepackage{amsbsy,amssymb}
\usepackage{amsmath,wasysym}
\usepackage{lineno}
\usepackage{lipsum}
\usepackage{epsfig}
\usepackage{color}
\usepackage{multirow}
\usepackage{textcomp}

\begin{document}

\title{Finite Size Effect in Amorphous Indium oxide }

\author{Sreemanta Mitra$^{\ddag}$}
\email[]{sreemanta85@gmail.com}
\author{Girish C Tewari}
\author{Diana Mahalu}
\author{Dan Shahar}
\email {dan.shahar@weizmann.ac.il}

\affiliation{Department of Condensed Matter Physics, The Weizmann Institute of Science, 76100 Rehovot, Israel}
\altaffiliation{Current address: Division of Physics and Applied Physics, School of Physical and Mathematical Sciences, Nanyang Technological University, Singapore 637371.}
\begin{abstract}
We study the low temperature magneto-transport properties of several 
highly disordered amorphous Indium oxide(a:InO) samples. Simultaneously
fabricated devices comprising a 2-dimensional (2D) film and 10 $\mu$m 
long wires of different widths were measured to investigate the 
effect of size as we approach the 1D limit, which is around 4 times the correlation length, 
and happens to be around 100
nm for a:InO. The film and the wires 
showed magnetic field ({\it B}) induced superconductor to insulator
transition (SIT). In the superconducting side, the resistance increased
with decrease in wire width, whereas, an opposite trend is observed 
in the insulating side. We find that this effect can be explained
in light of charge-vortex duality picture of the SIT.
Resistance of the 2D film follows an activated behavior over 
the temperature ($T$), whereas, the wires show a crossover from 
the high-$T$ activated to a $T$-independent behavior. 
At high temperature regime the wires' resistance follow the film's until
they deviate and became independent of $T$. We find that temperature at 
which this deviation occurs evolve with magnetic field and the width of the wire, 
which show the effect of finite size on the transport.

\end{abstract}
\maketitle
\section{Introduction}
Quantum phase transition (QPT) \cite{sondhirmp,sachdevbook}, 
an important paradigm of condensed matter physics, continues to generate immense experimental and  theoretical interest.
These transitions are accompanied by quantum rather than thermal fluctuations - where change in a parameter in the Hamiltonian
of the system induces a transition from one ground state to another fundamentally different ground state \cite{pu53}. 
Superconductor to insulator transition (SIT) is an archetypal QPT, which can be tuned by disorder \cite{haviprl62},
electric field \cite{goldmanprl94} 
or magnetic field ({\it B}) \cite{HebardPrl}. Although defined strictly at the absolute zero of temperature ({\it T}=0),
SIT has dramatic manifestations at experimentally accessible {\it T}s. 
SIT has been studied in several two-dimensional (2D) systems \cite{sit2}, from highly disordered superconductors \cite{Shaharprb,kapitulnikprl74,BaturinaJETP}
to high-$T_{C}$ superconductors \cite{bollnat}. Despite difficulties in fabrication, SIT has also been studied 
in several quasi-1D systems \cite{prl71,prb77,prl97,pr464}. Amorphous InO was also studied in quasi 1D form \cite{prl95} with width 
ranging from 40 to 100 nm. Some of the wires showed vanishing resistance as {\it T} is approached to zero below superconducting 
critical {\it T} ($T_{c}$)
and some saturated at a non-zero value ranging from 0.06-40 k$\Omega$. Below $T_{c}$, the magnetoresistance showed reproducible 
oscillations, which are reminiscent to that of superconducting interference device.   
\newline
The existence of superconductivity in the high disorder limit becomes more intriguing when the applied {\it B} induces vortices,
which remain mobile at any finite {\it T} due to the presence of disorder and a true superconducting state, where the vortices 
are localized can only be achieved at {\it T}=0. 
\par 
One of the central issues in the 2D SIT, is the nature of conduction in the insulating state of {\it B} tuned SIT,
where at low-{\it T} resistance per square ($R_{\Square}$)
become very high \cite{murthyprl1,murthyprl2}.
In several 2D disordered superconductors, such as thin MoGe films\cite{prl82}, Ta films\cite{prb73}, amorphous Bi films\cite{prl109}
($R_{\Square}$) become independent of {\it T} as the lowest {\it T} is approached. It was thought that formation 
of metallic state or macroscopic quantum mechanical tunneling of the vortices lead to this 
{\it T} independence of $R_{\Square}$ \cite{kapitulnikprl74,aliprl,prl82}. Weakening of the cooling of charge carriers
by the phonon bath due to weak electron-phonon coupling can also lead to {\it T} independent $R_{\Square}$.
Microwave spectroscopy on amorphous Indium oxide films \cite{prl111} provides evidence of {\it B}-induced metallic state
in the superconductor to insulator transition, where a crossover from superconductor to metallic state occurs at much
lower {\it B} than the conventional crossover. 
In case of going from 2D to quasi-1D system, several new features were observed. 
$T_{c}$ is reduced in case of amorphous MoGe films\cite{prl59} 
as the width is reduced and superconductivity was completely destroyed in the 1D limit. 
A recent work suggested that $T_{c}$ decreases exponentially with the inverse of wires' cross sectional area \cite{kimprl109}.  
The wires studied here are not 1D in strict sense, as per dimension is concerned, our only aim here is to see how the finite 
size affects the transport properties as we move down to the 1D limit. 
 Intuitively we expect that, if we reduce the width, $R_{\Square}$ should increase.
The as-prepared insulating wires' $R_{\Square}$ increase
 with reduction of width ({\it w}), however, 
a counterintuitive trend is observed in the insulating side for
the wires going through a {\it B}-driven SIT. In the insulator side of SIT, wires'$R_{\Square}$ increase with
{\it w}. 
\par 
In this Paper, we aim to address this puzzle by a detailed systematic size dependent 
investigation of magnetotransport measurements in a:InO devices. 
\section{Experiment}
The results which we present here are from 10 $\mu$m long wires with three different {\it w}, 0.1, 0.2 and 0.4 $\mu$m. 
The devices were initially defined by e-beam lithography on a Si/SiO$_{2}$ substrate and then 25 nm a:InO was 
electron-gun evaporated from ultrahigh pure sintered In$_{2}$O$_{3}$ targets in an oxygen atmosphere of residual
pressure 1.5 e-5 Torr. Apart from these wire devices, we prepare a film of size 50$\times$165 $\mu$m, simultaneously 
under the same condition to observe the finite size effect on electronic transport, for a:InO, as we reduce the width. 
All devices, wires and film, were patterned on a single chip and fabricated simultaneously.
The patterns were transfered using  lift-off process.  
\begin{figure}
\includegraphics[width=0.9\linewidth]{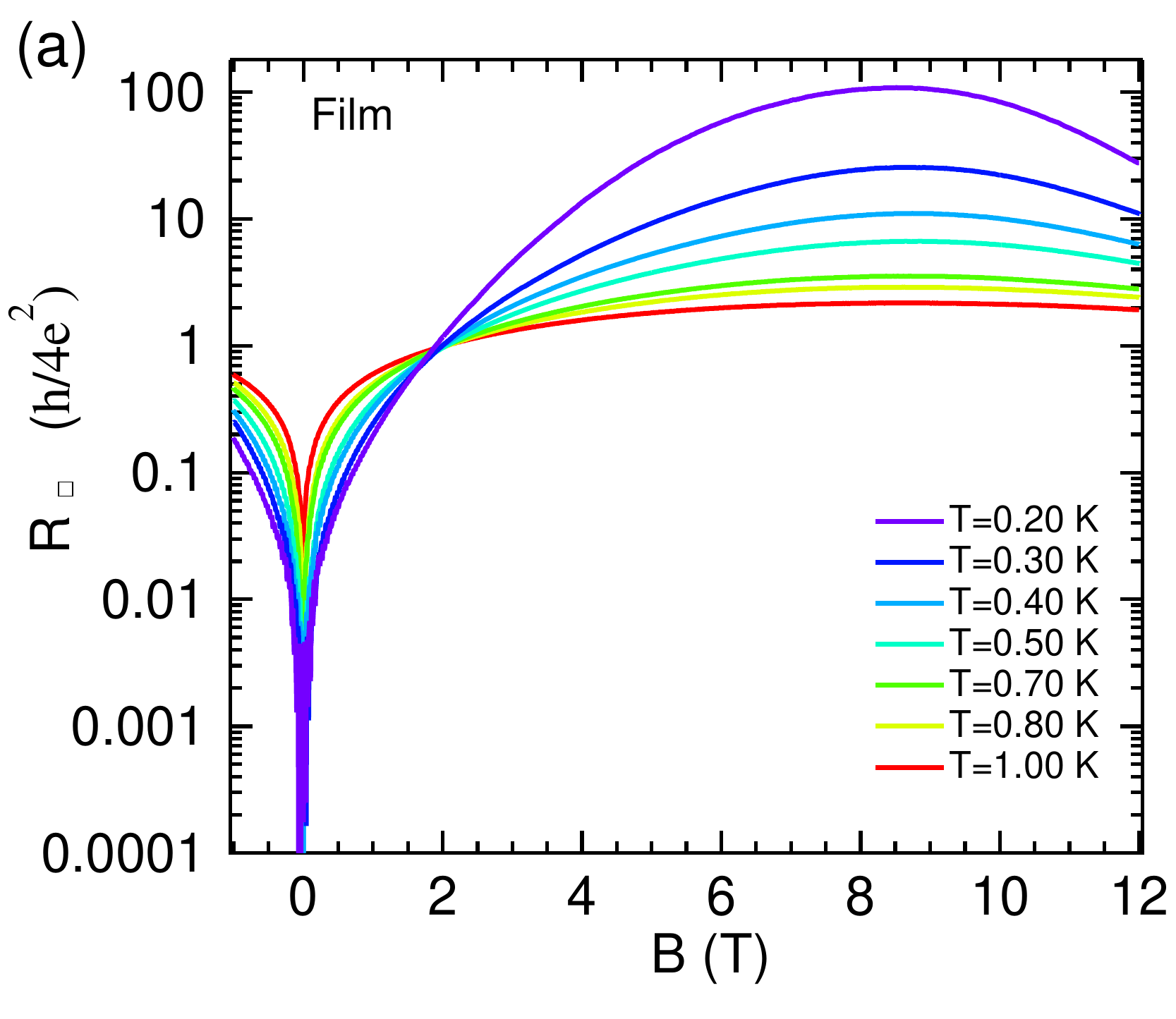}
\hspace{3pt}
\includegraphics[width=0.9\linewidth]{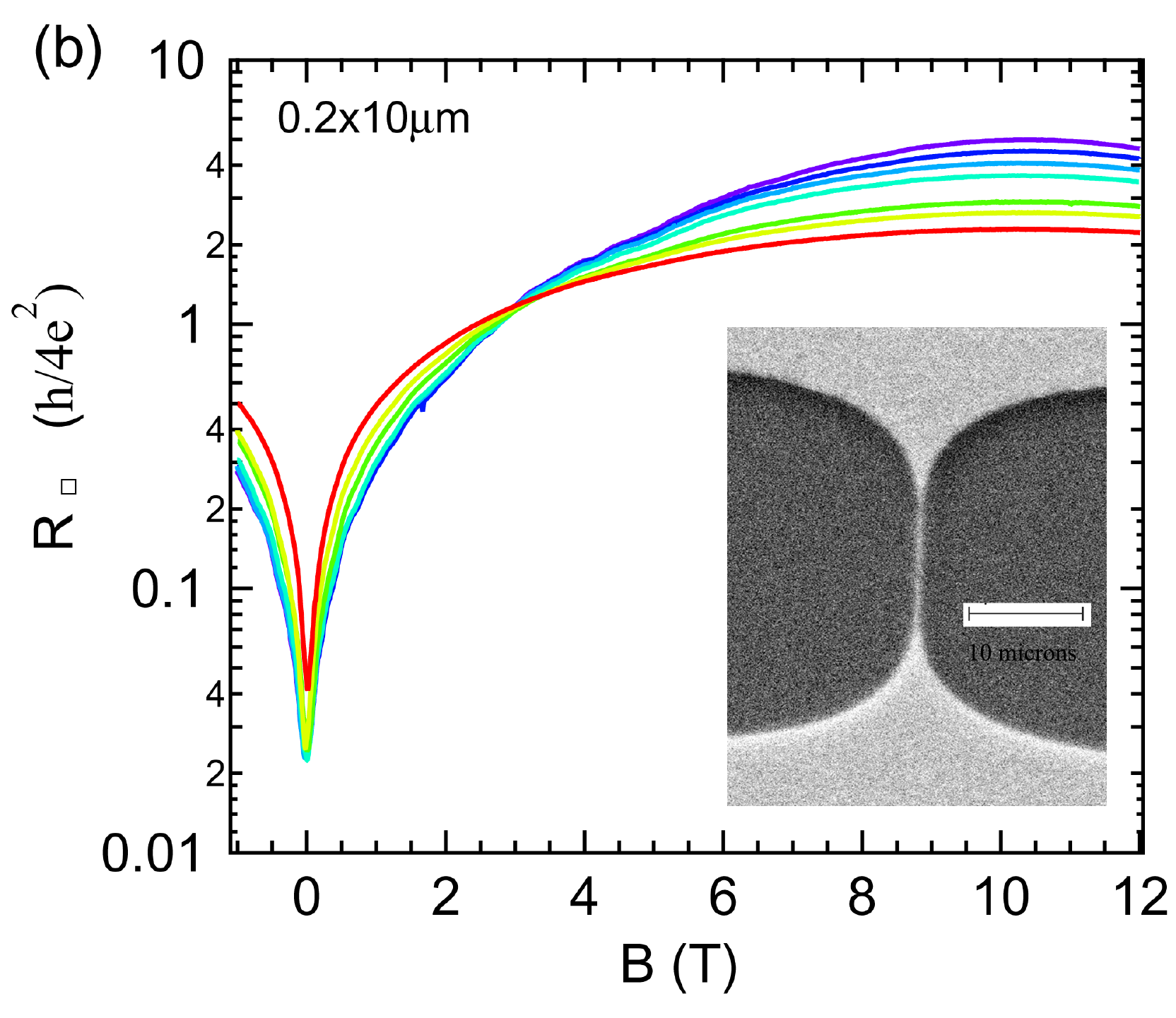}
\caption{(a) The {\it $R_{\Square}$ vs B} isotherms (in semi-log scale) for the film,
showing {\it B}-driven SIT ($B_{c}\cong $2.07 T). The $R_{\Square}$ scale is shown in the unit of 
quantum resistance of Cooper pair ($R_{Q}$=h/(2e)$^{2}$). The critical resistance ($R_{c}$) 
is almost equal to $R_{Q}$. 
 (b) The $R_{\Square}(B)$ isotherms for {\it w}=0.2 $\mu$m wire in a semi-log plot. 
 The critical magnetic field, $B_{c}$ is 3.03 T. At $B_{c}$, the resistance is 1.18$R_{Q}$.
 Same color legends for same {\it T} as (a) is used.
 (Inset: Scanning electron microscopy image of a 10 $mu$m long wire.)}       
\label{figrb1}
\end{figure}
For transport measurements, gold contact pads were
prepared by optical lithography followed by thermal evaporation and lift-off. The wires and the film of a:InO form a 
bridge between the contact pads. The chip was mounted on a chip carrier, and 25{$\mu$m} Au wires were 
used to bond for electrical connections between the devices and chip carrier gold pad. 
The inset of Figure \ref{figrb1}(b) shows the scanning electron microscopy image of a wire, used for the transport measurements.     
\newline 
The devices were cooled in a dilution refrigerator and magneto-transport measurements were performed using standard low frequency ac
lock-in technique in two-terminal configuration. The measurements were done in a current bias condition, 
with an applied bias $I_{ac}$=1 nA, which is well within the Ohmic regime.
The signal from the sample was amplified by a home made differential pre-amplifier and measured using lock-in amplifiers.
All the devices studied in the present work, a particular trend in the experimental result is observed and it’s not random in anyway. The magneto-transport isotherms were also found to be in appropriate way as observed previously in numerous occasions and follow the theoretical arguments perfectly. This systematics of the data suggested that heating or external noise has no significant effect on the measurements.
Magnetic field up to 12 T was applied perpendicular to the surface of the devices.  
Resistance mentioned here is the resistance per square, $R_{\square}$.
At room-{\it T}, and {\it B}=0 T, the film's $R_{\Square}$ is $\sim$1.4 k$\Omega$. 
The wires' $R_{\square}$, on the other hand, varies between 2.6-2.8 k$\Omega$. 
Prior to determine the $R_{\Square}$, the contact resistance ($\sim $490 $\Omega$) was subtracted. 
The subtracted resistance value was then normalized with the number of squares in series ({\it e.g.} 0.1 $\mu$m wide wire has 100
squares) to get $R_{\Square}$. The contacts are identical for all the devices, and most of the contact resistance  
originated from the connecting wires on the probe.  
\section{Results and Discussion}
We begin to present our findings by showing $R_{\Square}(B)$ isotherms obtained from our devices. 
Figure \ref{figrb1}(a) and (b) show $R_{\Square}(B,T)$ of the film and 0.2 $\mu$m wide wire respectively. 
The $R_{\Square}$ values are given in the unit of quantum resistance of
Cooper pair ($R_{Q}$=h/(2e)$^{2}$=6.45 k$\Omega$). Similar data have been obtained from the wires with {\it w}=0.1 and 0.4 $\mu$m. 
The isotherms as shown in Figure \ref{figrb1}, cross each other at particular {\it B}, $B_{c}$, signifying $B$-driven SIT.
The critical $R_{\Square}$, ($R_{\Square}$ value at $B_{c}$) $R_{c}$, for the film is close to $R_{Q}$ (1$\pm$0.05 $R_{Q}$), 
which suggests a phase transition driven by quantum phase fluctuations
and Cooper pair (de)localization expected within the bosonic description of SIT \cite{fisherprl}. 
For the wires, $R_{c}$ is not very different from $R_{Q}$; {\it e.g.} $R_{c}$=1.18$R_{Q}$ for {\it w}=0.2 $\mu$m wire. 
The values of the parameters for all the devices studied are summarized in Table \ref{tab1}. 
\begin{table}[!h]
\centering
\caption{List of parameters for all the devices. The $R_{\Square}$ values are given in terms of quantum resistance
of Cooper pair ($R_{Q}$=h/(2e)$^{2}$=6.45 k$\Omega$). {\it RT} stands for room temperature.}
\label{tab1}
\begin{tabular*}{8.5 cm}{lcccll}
\hline
\hline 
\multirow{2}{*}{Device}&$R_{\Square}$&\multirow{2}{*}{$B_{c} (T)$}&\multirow{2}{*}{$R_{c}$}&$R_{\Square}^{peak}$&$R_{\Square}^{peak}$\\
&      $(RT)$                 &           &       &  (1.0 K)&(0.2 K)\\
\hline 
2D Film & 0.217$R_{Q}$ & 2.07  & 1.0$\pm$0.05$R_{Q}$ & 2.17$R_{Q}$ & 107.8$R_{Q}$ \\
0.40 $\mu$m wire & 0.434$R_{Q}$ & 7.20  & 1.58$\pm$0.07$R_{Q}$ & 1.98$R_{Q}$ & 3.68$R_{Q}$ \\
0.20 $\mu$m wire & 0.403$R_{Q}$ & 3.03  & 1.18$\pm$0.08$R_{Q}$ & 2.28$R_{Q}$ & 4.98$R_{Q}$ \\
0.10 $\mu$m wire & 0.418$R_{Q}$ & 1.30  & 0.92$\pm$0.06$R_{Q}$ & 2.83$R_{Q}$ & 6.73$R_{Q}$ 

\end{tabular*}
\end{table}
 
 To look at the nature of transport on both sides of the SIT, we plot the {\it T} dependence of $R_{\Square}$ for various {\it B} values. 
 The data were extracted from the $R_{\Square}(B)$ isotherms obtained for the Film and the 0.2 $\mu$m wire as
 shown earlier in Figure \ref{figrb1}.   
 \begin{figure} [!h]
 	\includegraphics[width=0.9\linewidth]{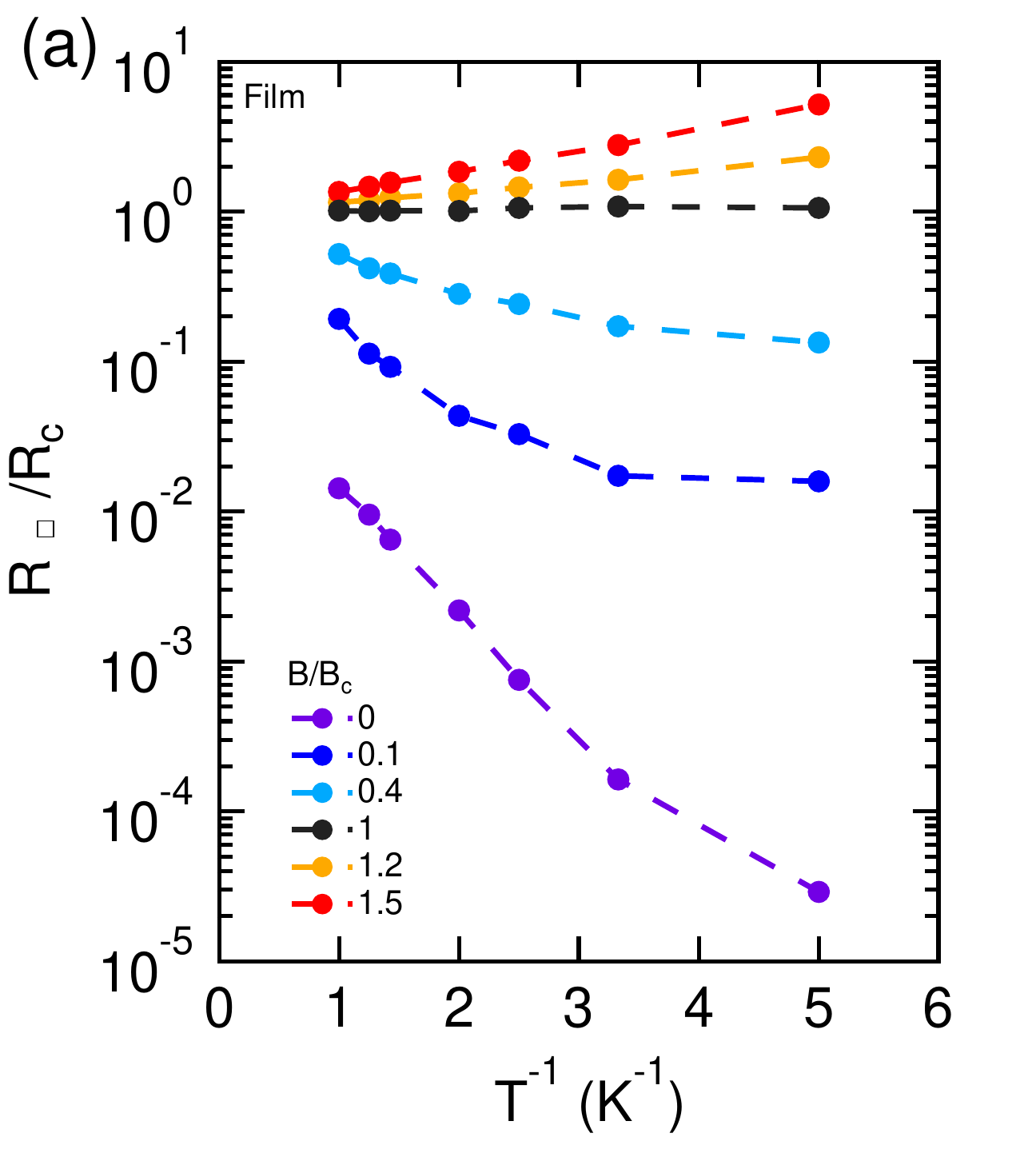}
 		\includegraphics[width=0.9\linewidth]{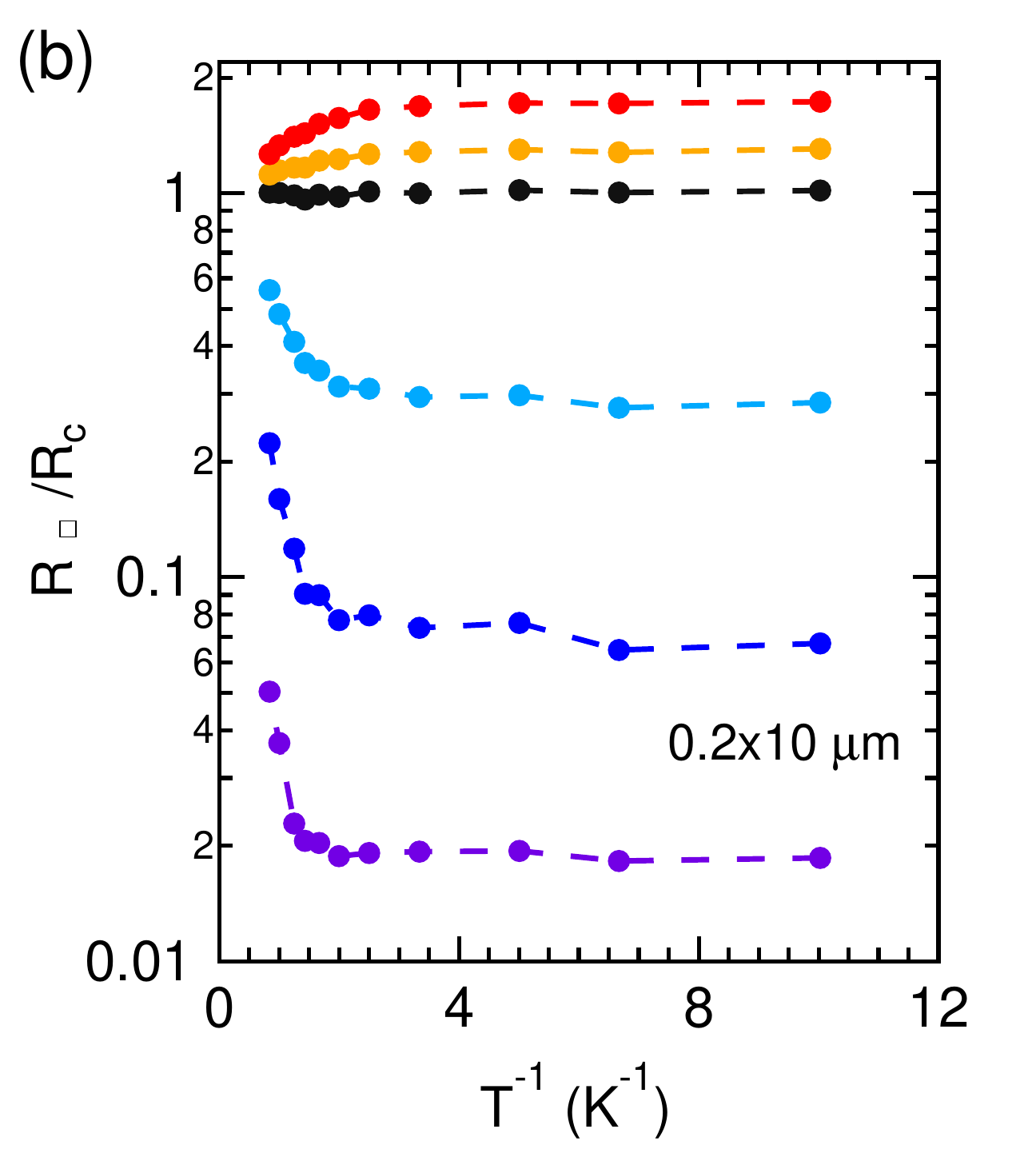}
 	\caption{(a) Arrhenius plot of $R_{\Square}$ (normalized with respect to $R_{c}$) of the film for six 
 		different ($\frac{B}{B_{c}}$) values. (b)  Similar plot of $R_{\Square}$ of the $w$=0.2 $\mu$m wire for
 		the same ($\frac{B}{B_{c}}$) values as (a). Same color scheme has been used in both the figures. 
 		The dashed lines joining the data points are to guide the eye. The wire's $R_{\Square}$ show a crossover
 		from the high-{\it T} activated to a {\it T} independent behavior as {\it T} is lowered.}       
 	\label{figrt1}
 \end{figure}
 Figure \ref{figrt1}(a) shows an Arrhenius plot of $R_{\Square}$ for 6 different ($\frac{B}{B_{c}}$) values ranging from
 the superconducting to the insulating state for the film.
 At {\it B}=0, $(R_{\Square})_{Film}$ decreases and appeared to go to zero as {\it T} is decreased, signifying superconducting ground state. 
 For the wires, on the other hand, $R_{\Square}$ saturate as {\it T} is decreased. 
  For ($\frac{B}{B_{c}}<$1), $R_{\Square}$ drops as {\it T} is decreased, 
 following,
 \begin{equation}
 R=R_{0}exp(-\frac{T_{0}}{k_{B}T})
 \label{eq1}
 \end{equation}
 where, $T_{0}$ is the activation energy, $k_{B}$ is the Boltzmann constant and $R_{0}$ is a pre exponential normalization factor.
 Increasing the magnetic field, increases the low-{\it T} limit of $R_{\Square}$. 
 In the superconducting side, for nonzero {\it B}, $R_{\Square}$ deviated from the high-{\it T} 
 activated behavior as {\it T} is lowered and tend to become independent of {\it T}. 
 It was thought that, near SIT, disordered superconducting films may intrinsically break up into islands, 
 which might be the reason behind this {\it T} independence of $R_{\Square}$ at low-{\it T}s\cite{prb65,nat449}. 
 In the insulating side, above $B_{c}$, no such deviation is observed and $R_{\Square}$ increases rapidly 
 as a function of $(1/T)$ over the full dynamic range. 
 This behavior can be fitted with the activated form described in Equation \ref{eq1}.
 In Figure \ref{figrt1}(b), 
 we show similar Arrhenius plots for the 0.2 $\mu$m wide wire, for the same ($\frac{B}{B_{c}}$) values as
 in Figure. \ref{figrt1}(a). The data clearly show a crossover from the high-{\it T} activated behavior to a 
 leveling of $R_{\Square}$ at low-{\it T}s.
 Similar behavior has been observed for the other wires studied here. At zero-{\it B}, this leveling
 might be suggestive of phase-slip processes \cite{giordanoprl}. At non-zero {\it B}, this crossover
 to a {\it T}-independent state might indicate that the system change from  activated to a metallic dissipative 
 state \cite{aliprl}.
 Another effect that would cause this tendency of the resistance 
 to flatten at low-{\it T} is thermal decoupling of the electrons from the phonon bath, but this is more 
 likely to occur at much lower {\it T}\cite{maozprl,maozsrep} than 0.8 K, where it is actually seen in this study. 
 It was argued in ref. \cite{aliprl} 
 that this leveling of resistance at low {\it T} possibly originates from a macroscopic quantum mechanical 
 tunneling of vortices between pinning sites, where the tunneling objects are dislocations and antidislocations 
 of the vortex lattice. Also a model \cite{spivakprl,jimprb} for crossover from a vortex solid phase to a quantum
 vortex liquid phase suggests temperature independent $R_{\Square}$ at low-{\it T}. 
\begin{figure}
	\includegraphics[width=.95\linewidth]{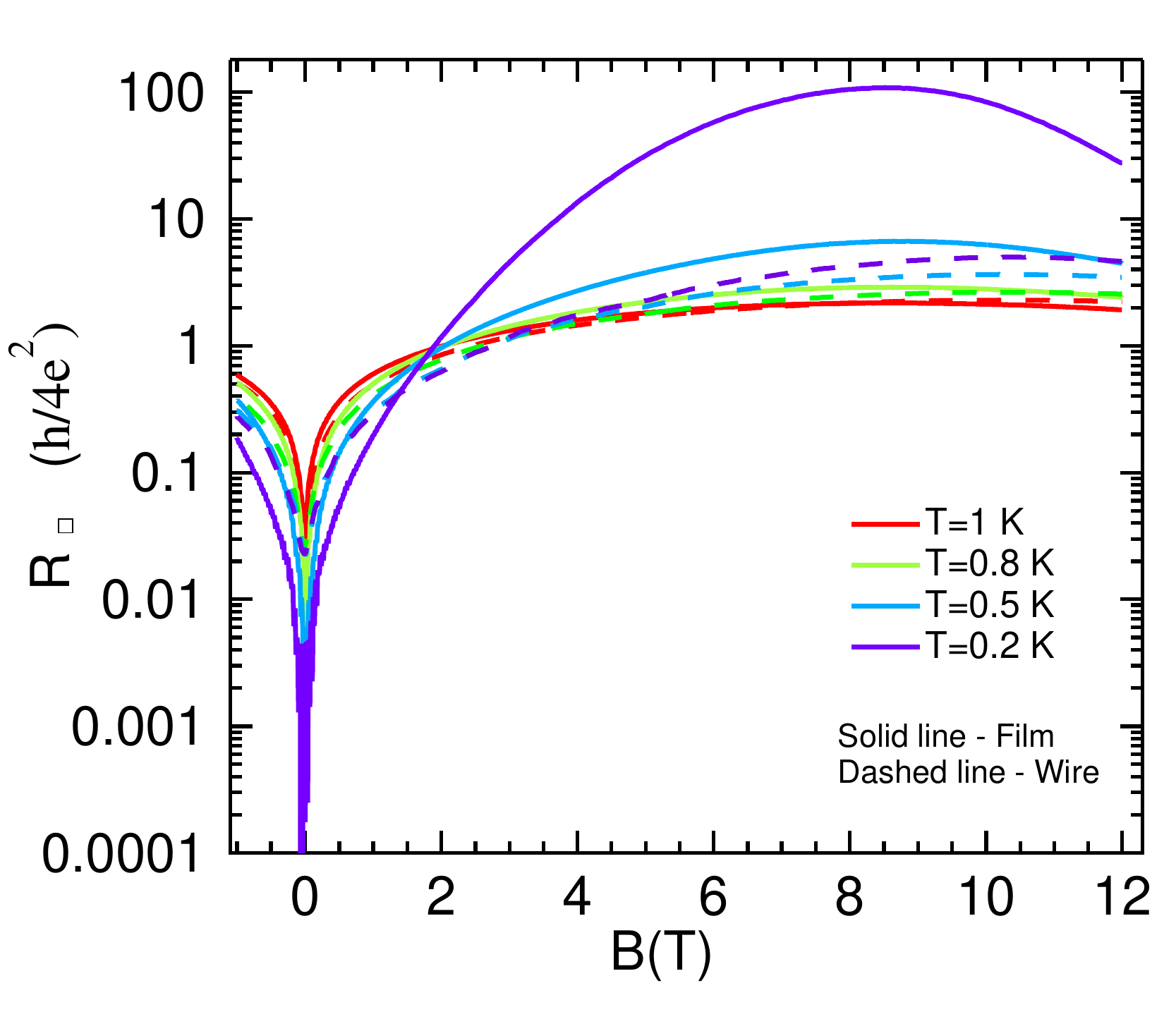}
	\caption{$R_{\Square}(B)$ isotherms for both film and 0.2 $\mu$m wire, at 4 different {\it T}s. 
		Same color has been used for same temperature. The solid and dashed lines are for the film and the wire respectively. 
		For the lowest {\it T} shown, $(R_{\Square})_{Film}$ is order of magnitude higher than $(R_{\Square})_{wire}$, in the 
		insulating side.}       
	\label{figrb2n}
\end{figure}
\begin{figure} 
\includegraphics[width=1.05\linewidth]{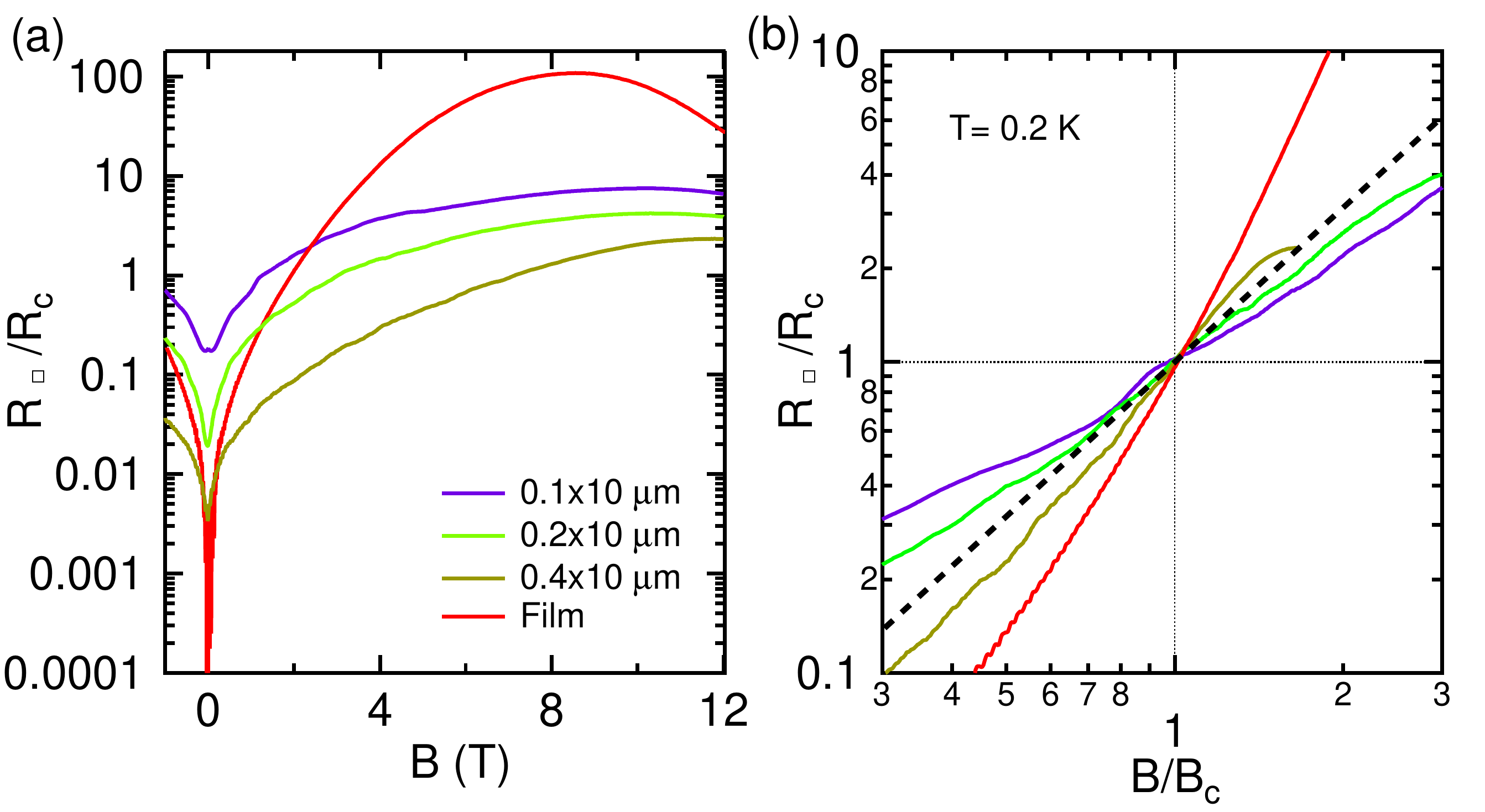}
\caption{(a) $R_{\Square}(B)$ for all the wires along with the film measured at {\it T}=0.2 K. 
The resistance values are normalized with respect to 
respective $R_{c}$ of the devices. 
(b) The $R_{\Square}(B)$ for the same devices measured at 0.2 K, shown in double-logarithmic plot.
The {\it B} scale is normalized with respect to respective $B_{c}$ values. The linearity in double 
logarithmic plot near ($\frac{B}{B_{c}}$=1), suggests power law behavior. A dashed straight line is drawn to guide the eye. 
The linear behavior extends deep in to the insulating side, suggests that vortex transport play an important role in the insulating regime.}       
\label{figrb2}
\end{figure}
Although the isotherms for the film and the 0.2 $\mu$m wide wire look similar in appearance, and $R_{\Square}$
increases with {\it B} until it reaches a magnetoresistance peak in the insulating side,
there are significant differences.   
In the superconducting side of the SIT, the wire has higher $R_{\Square}$ than the film. 
At high-{\it T},  $R_{\Square}$ of the film is of the same order of magnitude as wire [see 
Figure \ref{figrb2n}].
However, the film's $R_{\Square}$ increases faster with {\it B}, giving rise to order of magnitude higher $R_{\Square}$ in the insulating state at low{\it T}.  
\par        
This is seen more clearly when we plot in Figure \ref{figrb2}(a) the $R_{\Square}(B)$ for the wires with different {\it w},
along with the film, measured at {\it T}=0.2 K. Since $R_{c}$ changes with the width of the nanowires (see Table \ref{tab1}), 
we normalize $R_{\Square}$ with respect to the $R_{c}$ values for comparison. At {\it B}=0,
the film has lowest $R_{\Square}/R_{c}$, and $R_{\Square}/R_{c}(B=0)$ increases gradually as we decrease the width of the wires. 
This changes when we re-plot the data in Figure \ref{figrb2}(a) with a {\it B}-scale normalized with respect 
to respective $B_{c}$, in Figure \ref{figrb2}(b). On both sides of $B_{c}$, $R_{\Square}$ follow a
trend with size; below $B_{c}$, film has smallest $R_{\Square}$, and it increases with the reduction in the width of nanowire, 
whereas for $B>B_{c}$, the film has more resistance and it decreases as {\it w} is reduced progressively. 
\newline 
The apparent linearity in the double-logarithmic representation of data as shown in Figure \ref{figrb2}(b), 
for the wires along with the film, emphasize that the
wires follow a similar kind of phenomenological power law behavior \cite{murthyepl,maoznat}.
In the superconducting side
of the {\it B}-tuned SIT, the power law is consistent with the collective pinning 
model of transport \cite{rmp66,aliprl,murthyepl} 
and 
indicate the role played by interacting bosons (vortices) in transport.
The existence of the power law behavior above $B_{c}$, 
suggest that vortex transport might be important in the insulating state \cite{murthyepl,maoznat}. 
\par 
A quantum critical point is known to separate ground states with different symmetry. Being a QPT, criticality in SIT, results in scaling laws \cite{sondhirmp,HebardPrl}. These scaling relation was obtained using the characteristic length scales, set either by {\it T} or 
electric field ({\it E}) \cite{kapitulnikprl74}. It reduces to a single variable function if either of {\it T} or {\it E} is fixed. The {\it T}-scaling of $R_{\Square}$ provides the exponent $z\nu$, since the abscissa is scaled as 
$u=| B-B_{c} | / T^{z\nu}$. As a result of this {\it T}-scaling, all the curves have collapsed into one, showing at
{\it T} =0 a bifurcation, or unstable fixed point. \newline For each device, $R_{\Square}(B, T, E=0)$ data were used to test the scaling prediction, by varying {\it B} at small increments at fixed {\it T}. 
\begin{figure} 
	\includegraphics[width=1.05\linewidth]{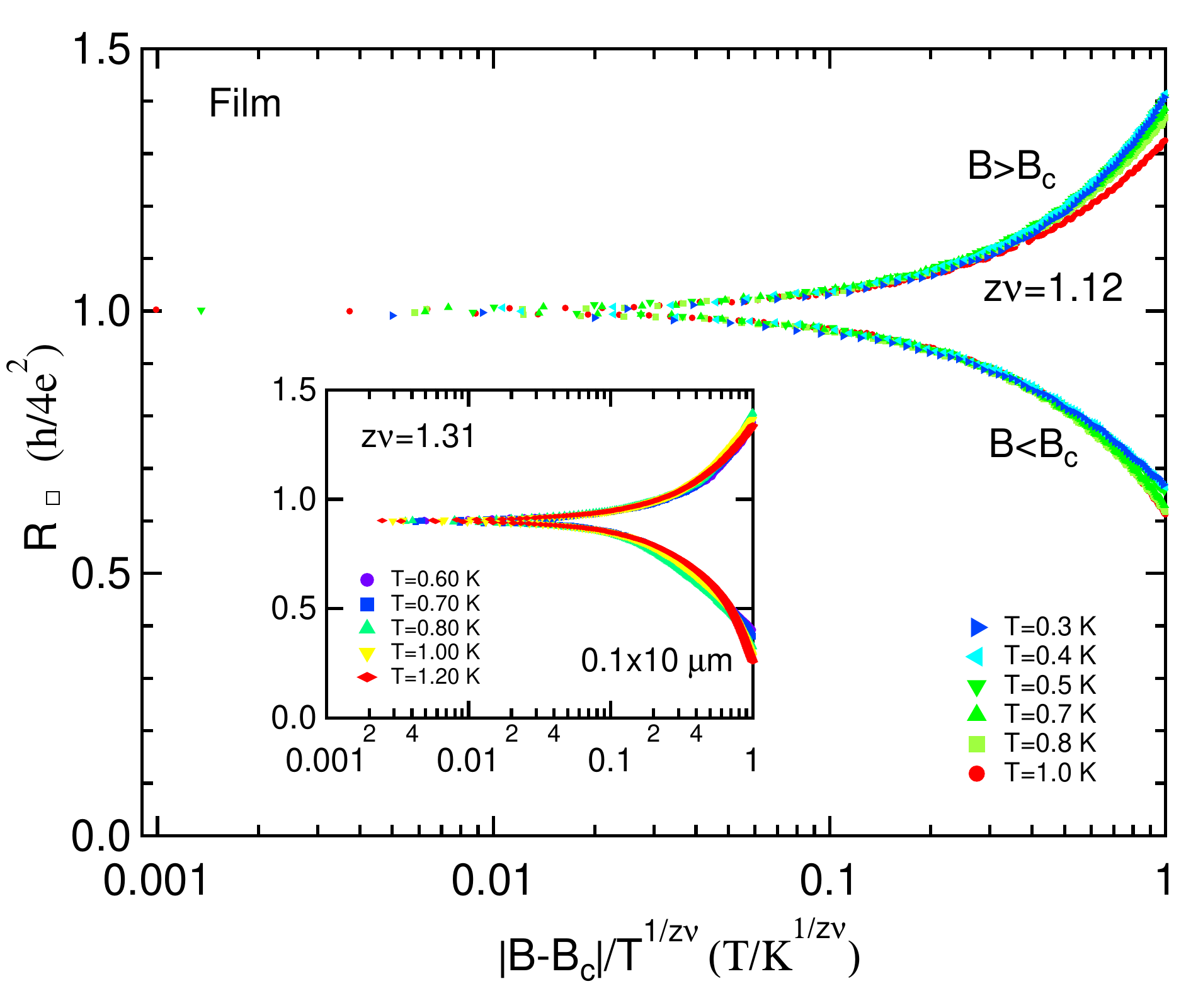}
	\caption{Temperature scaling for the Film and 0.1 $\mu$m wide wire (inset). The shown exponent $z\nu=1.12$ and ($z\nu$=1.31) is determined from the best visual data collapse. }       
	\label{figsca}
\end{figure}
In figure \ref{figsca}, the {\it T}-scaling of $R_{\Square}$ for the film is shown. A good visual collapse of the data is obtained for $z\nu$=1.12, which is close to the previous findings\cite{HebardPrl,maoznat}. For the wires, $R_{\Square}$ flattens and become {\it T} independent on either side of $B_{c}$ [figure\ref{figrt1}(b)], as low-{\it T} is approached. For obvious reason this affects the scaling. Hence, we prefer to scale wires' $R_{\Square}$ only for $T\geq$ 0.6 K. In the inset of figure \ref{figsca}, the scaling results 
for one of our wire devices (0.1$\times$10 $\mu$ m) is shown. For the best visual collapse was obtained for $z\nu=1.31$ which is not very different from that observed for the film. The scaling analysis near the critical point allows one to get the symmetry classification without the functional form the data follow\cite{maoznat}. Although in our system the power law as shown in Figure \ref{figrb2}, can be treated as entire scaling function.      
\par
The high-{\it T} activated behavior of $R_{\Square}$ for all the devices allows us to extract the activation energies ($T_{0}$). 
We find that $T_{0}$ varies logarithmically with {\it B}, following $T_{0}\sim \ln (\frac{B}{B_{c}})$ as observed in our
previous study\cite{murthyepl}. Theoretically, this 
variation of $T_{0}$ is consistent with the collective pinning model of vortex transport \cite{rmp66,aliprl,murthyepl}. 
This type of logarithmic
field dependence of activation energy is also found in several other cases, such as thermally activated motion
of vortex-antivortex pairs \cite{epl20} or activation over surface barrier associated with edge pinning \cite{KoshphyC}.  
\begin{figure} 
	\includegraphics[width=\linewidth]{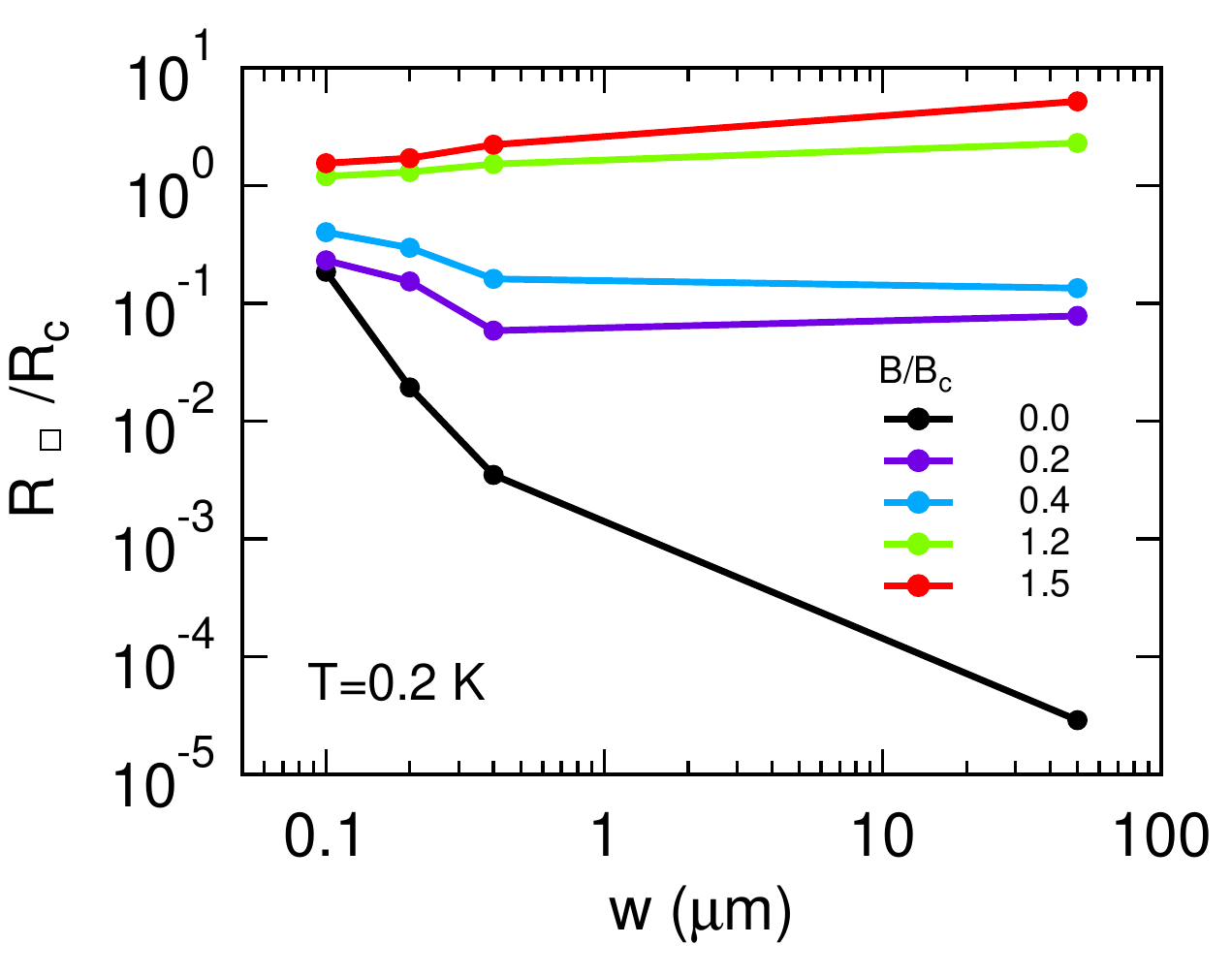}
	\caption{ The variation of $R_{\Square}$ with width ({\it w}) of wires at different ($\frac{B}{B_{c}}$) values measured at 
		{\it T}=0.2 K. 
		$R_{\Square}$ is normalized with respect to the respective $R_{c}$ values. A log-log representation is used to include the 50 $\mu$m wide film's data in the plot.
		In the superconducting side of the SIT, $R_{\Square}$ decreases as {\it w} is 
		increased, whereas, in the insulating side, $R_{\Square}$ increases with an increase in {\it w}.  }       
	\label{figrw}
\end{figure}
\par 
To show the dependence of $R_{\Square}$ on the width of the nanowires we present in Figure \ref{figrw}, $R_{\Square}(w)$ for different 
($\frac{B}{B_{c}}$) values measured at 0.2 K.  In the superconducting side of the SIT, $R_{\Square}$ decreases as {\it w} is 
increased, whereas, in the insulating side, $R_{\Square}$ increases with {\it w}. 
A log-log representation is used to include the film's data for the same ($\frac{B}{B_{c}}$) values as of the wires.
It is seen that the film's data follows the wires' trend as $w\rightarrow\infty$. 
\newline
\begin{figure}[!h]
\includegraphics[width=0.95\linewidth]{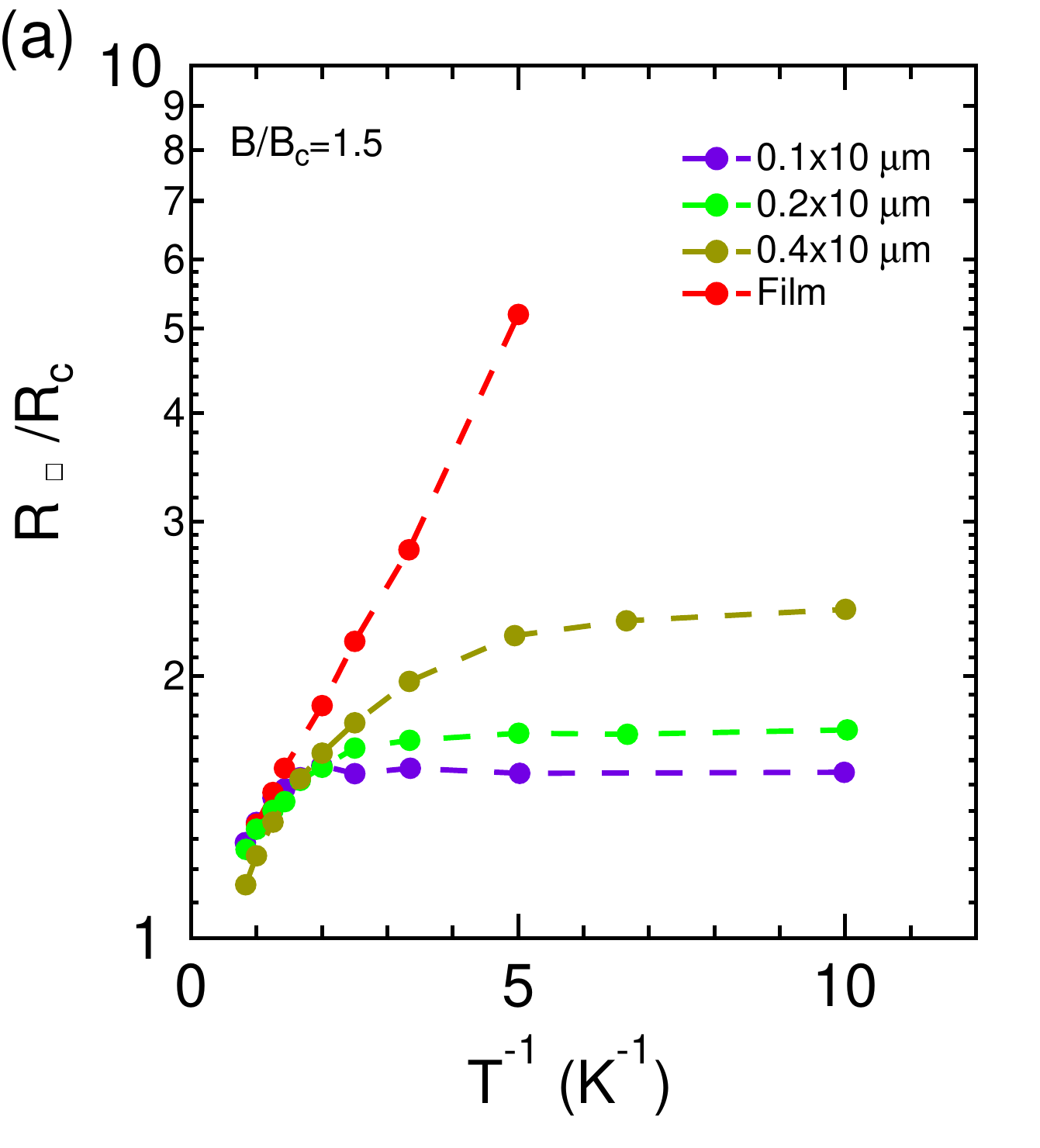}
\hspace{3pt}
\includegraphics[width=0.95\linewidth]{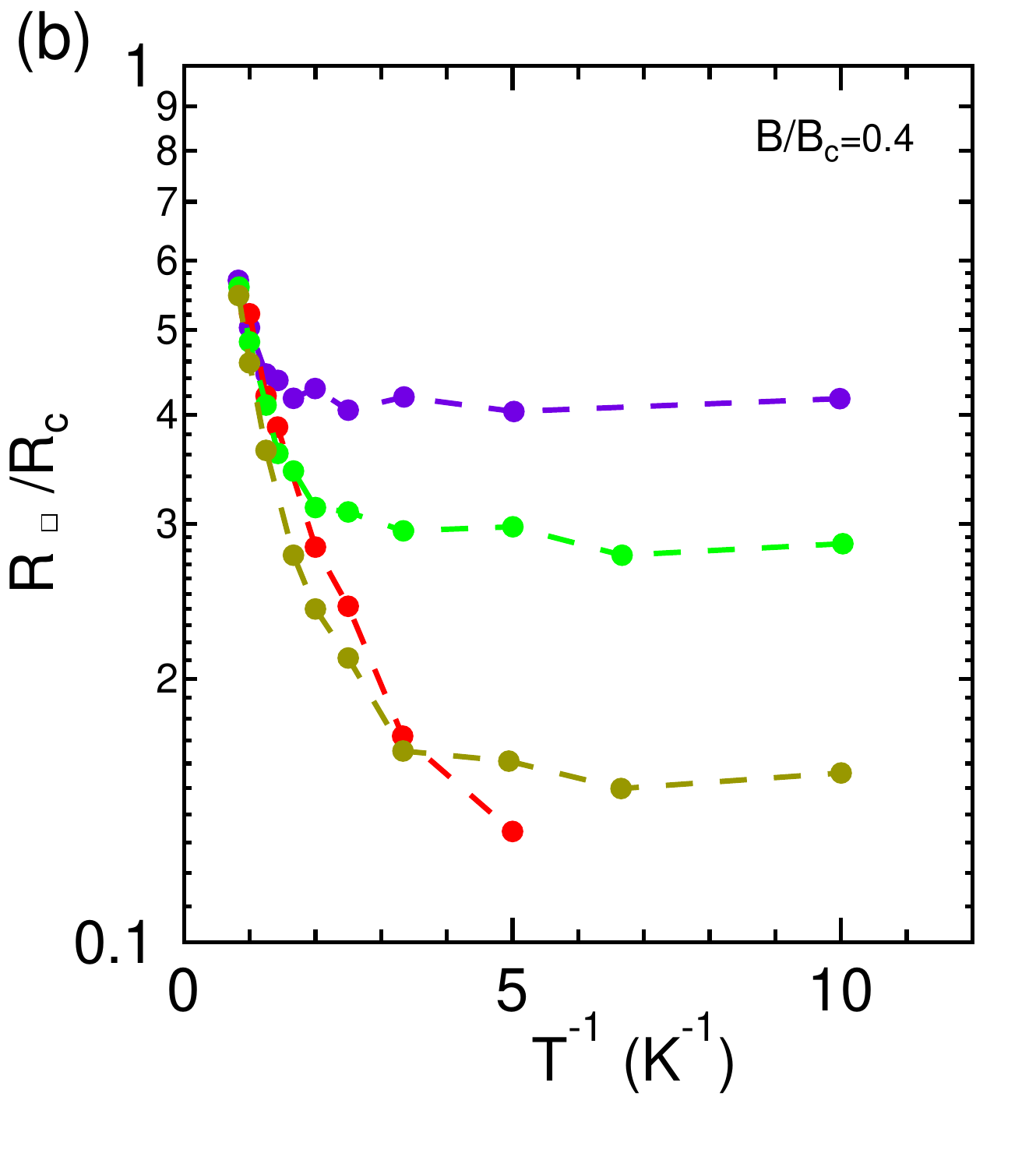}
\caption{Variation of normalized $R_{\Square}$ with $T^{-1}$ for wires of different widths along with the 
film measured at (a) ($\frac{B}{B_{c}}$=1.5) and (b) ($\frac{B}{B_{c}}$=0.4). 
At high-{\it T} all the devices follow the same activated behavior, until the wires' $R_{\Square}$ become independent
of {\it T}. Below $B_{c}$, the wire with lower {\it w} has more $R_{\Square}$ whereas, above $B_{c}$, $R_{\Square}$ increases 
as {\it w} is increased.}       
\label{figrt2}
\end{figure}
To illustrate the effect of size on the superconducting and insulating sides, Arrhenius plots of $R_{\Square}$, for all 
the devices at same ($\frac{B}{B_{c}}$) values is shown in Figure \ref{figrt2} (a) and (b). $R_{\Square}$ was normalized with
respect to respective $R_{c}$ values for  comparison. At high-{\it T}, the wires' $R_{\Square}$ follow the same activated
form as the film, until they deviate and become {\it T}-independent. The value of the saturated resistance, ($R_{sat}$), 
follows different trend with {\it w} on either side of the {\it B}-tuned SIT. In the superconducting side 
[as inferred from figure \ref{figrt2}(b)], $R_{sat}$ increases as {\it w} is reduced whereas in the insulating side
[Figure \ref{figrt2}(a)] $R_{sat}$ decreases with a decrease in {\it w}. 
\par 
This unique width dependence of resistance in the insulating state, can be explained 
if we adhere to the charge-vortex duality picture of the SIT. In the superconducting phase, Cooper pairs form the superfluid condensate, whereas, vortices are localized by forming bound vortex antivortex pairs, 
and their motion give rise to resistance. As the channel width is reduced, the vortices can tunnel or hop from one end to another and $R_{sat}$ is increased. On the other hand, in the insulating side, the vortex-antivortex pairs are broken by quantum fluctuations at low-{\it T}, whereas the Cooper pairs are localized and form Bose glass. 
The motion of these Copper pairs would increase the conductance (reduction in resistance) in the system. Upon the reduction of the channel width, the movement of the localized 
Cooper pairs give lower $R_{sat}$ in the narrower wires. This apparently leads to the fact that 
as we keep on decreasing the width, $R_{\Square}$ will be less in the insulating side at low-{\it T} and quasi-1D systems
would never show as high $R_{\Square}$ as of 2D films over the same $(\frac{B}{B_{c}})$ range. 
\par    
The temperature, $T_{D}$ where the wires' $R_{\Square}$ deviate from the film, evolve with {\it w} and {\it B}. 
\begin{figure} 
\includegraphics[width=0.98\linewidth]{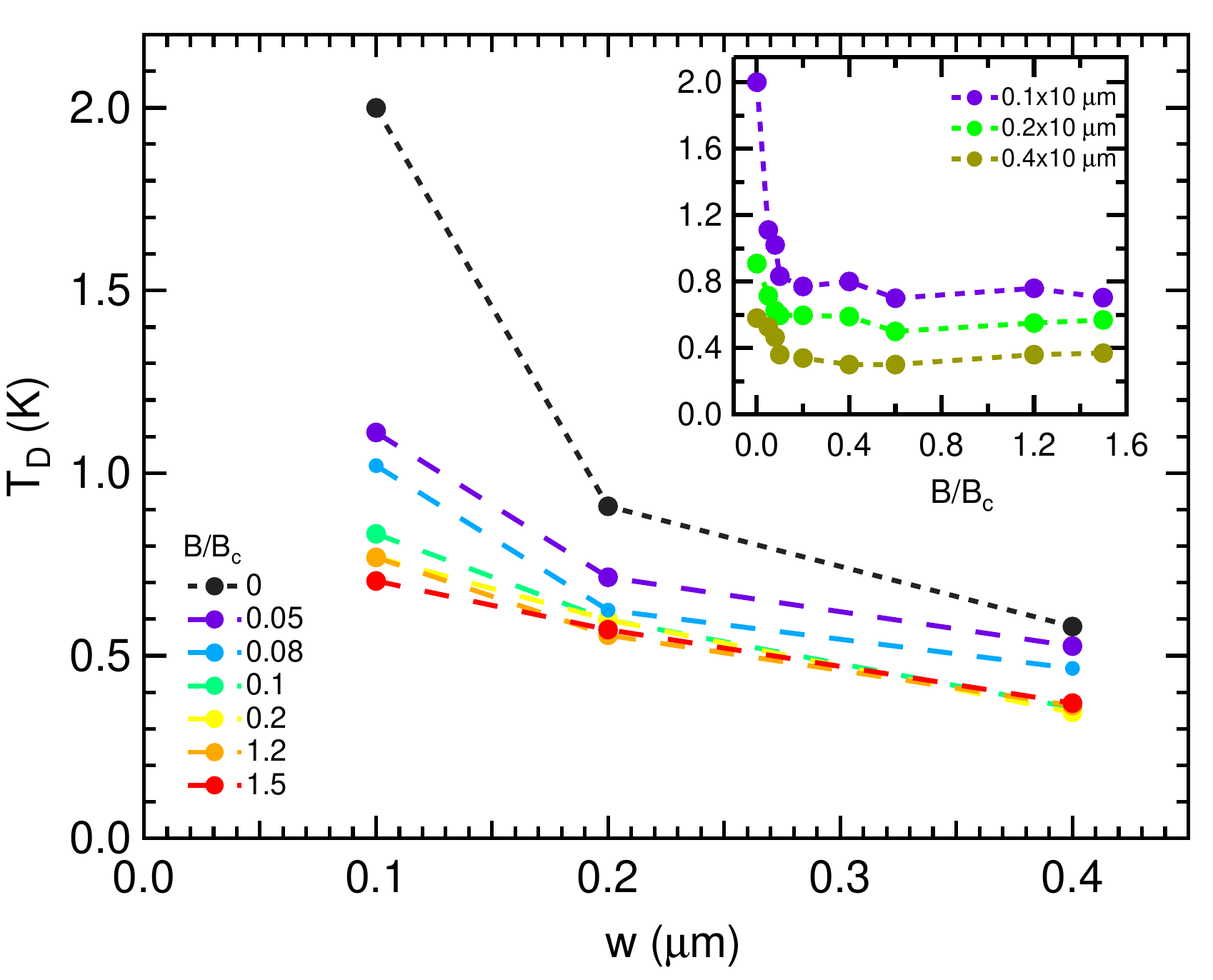}
\caption{Variation of temperature of deviation ($T_{D}$) (see text) with {\it w} for different ($\frac{B}{B_{c}}$) values. 
(Inset): Variation of $T_{D}$ with ($\frac{B}{B_{c}}$) for wires with different widths. 
For a particular ($\frac{B}{B_{c}}$), $T_{D}$ decreases with increase in {\it w}. 
For a particular {\it w}, $T_{D}$ initially decreases with increasing ($\frac{B}{B_{c}}$) and then saturate.}       
\label{figtd}
\end{figure}
In Figure \ref{figtd} the variation of $T_{D}$ with {\it w} for different ($\frac{B}{B_{c}}$) values of the wire is shown. 
For a particular ($\frac{B}{B_{c}}$), $T_{D}$ increases with decrease in {\it w}. For 0.4 $\mu$m wide wire, $T_{D}$ is smallest 
of the three, as it is more close to the film, as per as dimension is concerned in comparison to the 0.2 and 0.1 $\mu$m
 wide wire.
Also the evolution of $T_{D}$ is found to have a dependence on the magnetic field, in the superconducting 
side, ($\frac{B}{B_{c}}<$1). In the inset of Figure. \ref{figtd} the variation of $T_{D}$ with ($\frac{B}{B_{c}}$) is shown.
$T_{D}$ is maximum at zero magnetic field and decreases as {\it B} is increased in the superconducting 
side of the {\it B}-tuned SIT, and attained saturation, which continued above $B_{c}$. 
\section{Summary}
We performed magnetotransport measurements on several 10 $\mu$m long nanowires with different widths,
along with a 2D film of a:InO. The devices showed {\it B}-driven superconductor to insulator transition. 
On either side of the {\it B}-driven SIT, the resistance behave differently but systematically with width. 
In the superconducting side, $R_{\Square}$ increases with decrease in width, whereas in the insulating side, it behaves oppositely.
Initially, at high-{\it T} all of them has similar activated transport, however, as {\it T} is reduced the wires deviate from 
that and $R_{\Square}$ become independent of {\it T}. The temperature of deviation of wires' resistance from that of film, is found 
to evolve with the width of the wire and magnetic field. Our analysis suggest that at low-{\it T}, high resistance in the 
insulating side of SIT will be absent in narrower samples due to finite size effect on the motion of the localized Cooper pairs.   
\section{Acknowledgments} 
S.M. and G.C.T. thank Sumilan Banerjee for fruitful discussions and department's sub-micron center for providing necessary 
support in device fabrication. S.M. thanks VATAT program for post-doctoral fellowship support. This work
was supported by the Minerva Foundation with funding from Federal German Ministry for
Education and Research, and by a grant from the Israel Science Foundation.
%

\end{document}